\documentclass[a4paper,12pt]{article}

\usepackage[left=2.5cm,right=2.5cm,top=2.5cm,bottom=2.5cm]{geometry}
\usepackage{graphicx}
\usepackage{amssymb}
\usepackage{multirow}
\usepackage{amsmath}
\usepackage{psfrag}
\usepackage{setspace}
\usepackage{subcaption}
\usepackage[colorlinks=true,bookmarks=true]{hyperref}
\usepackage{float}
\usepackage{cite}
\usepackage[utf8]{inputenc}
\usepackage{authblk}

\usepackage{mathtools}

\DeclarePairedDelimiterX\braket[2]{\langle}{\rangle}{#1 \delimsize\vert #2}

\hypersetup{citecolor=blue}
\newcommand{\dif}{\mathrm{d}}
\newcommand{\Eqref}[1]{(\ref{#1})}
\newcommand{\half}{\frac{1}{2}}
\newcommand{\expo}[1]{\mathrm{e}^{#1}}
\newcommand{\brac}[1]{\left(#1 \right)}
\newcommand{\sbrac}[1]{\left[#1\right]}

\begin{document}

\title{Criteria of existence for bounce solutions in false vacuum decay with gravity}
\author{Nicholas W.~K.~Wong\footnote{whye\_khuin\_nicholas\_wong@moe.edu.sg} \\\vspace{-12pt}
 {\small \textit{Department of Physics, River Valley High School, 649961, Singapore} \\\vspace{-6pt} 
 \textit{Ministry of Education, 138675, Singapore}}
 \par \vspace{6pt}
 Jiangbin Gong\footnote{phygj@nus.edu.sg}, Yen-Kheng Lim\footnote{phylyk@nus.edu.sg}, and Qing-hai Wang\footnote{qhwang@nus.edu.sg} \vspace{-6pt}\\
 {\small \textit{Department of Physics, National University of Singapore, 117551, Singapore}}
}

\renewcommand\Authands{ and }

\date{\normalsize{\today}}
\maketitle

\begin{abstract}
  The bounce solutions of self-interacting scalar fields coupled to gravity are studied using a semi-classical approach. We found that bounce solutions have a \emph{maximum} required barrier curvature, in addition to the known \emph{minimum} required barrier curvature. In particular, as the maximum barrier curvature is approached, the scale factor of the well-known Colemen-De Luccia (CDL) bounce solutions become divergent. Unlike the CDL or its more general oscillating bounce counterparts, this cannot be considered as a subset of the Hawking-Turok solution. 
\end{abstract}

\section{Introduction} \label{intro}

The study of false vacuum decays in the presence of gravity \cite{Coleman:1980aw} plays an important role in describing quantum effects in the early universe. In these models, the false vacuum is an initial meta-stable state with a constant scalar field. The scalar potential at this state corresponds to an effective cosmological constant. Transitions between the two vacua may describe the phase transition between any combination of de Sitter and Anti-de Sitter geometries. The de Sitter case is of particular interest since it provides a useful model for the inflationary epoch in the early universe \cite{Guth:1980zm,Linde:1981mu,Albrecht:1982wi}.

The semi-classical description of false vacuum decays in flat space has been well understood for some time \cite{Kobzarev:1974cp,Frampton:1976kf,Coleman:1977py,Callan:1977pt}. A generalisation of the problem to include gravity was considered by Coleman and De Luccia (CDL) \cite{Coleman:1980aw}. In both flat- and curved-space descriptions, the decay rate is calculated from the Euclidean action evaluated under the classical path which satisfies the Euler-Lagrange equations. It follows that a full understanding of the vacuum decay requires knowledge of the classical solutions to the field equations. 

To model the decay of the field, the potential function for the field $\phi$ is required to have two minima separated by a barrier. Thus a natural choice that captures this essential features is a potential function quartic in $\phi$, which depicts an asymmetric double well. The classical bounce\footnote{Here, the term `bounce' refers to the oscillation of the scalar field, as opposed to a similar terminology used in the context of `Big Bounce' cosmology which refers to time-evolution of the scale factor in Friedmann-Robertson-Walker cosmology, for instance studies in Ref.~\cite{Odintsov:2015ynk,Nojiri:2017ncd}} solution under such a potential was found by CDL. It describes $\phi$ being initially at the false vaccuum, then crossing the barrier to reach the true vacuum. The original solution presented by CDL was obtained under the thin wall approximation, where the energies at the false and true vacua are infinitesimally close to each other. Since then, numerical solutions without the thin wall approximation have also been obtained \cite{Konoplich:1986zp,Isidori:2001bm,Strumia:1998qq,Baacke:2003uw,Dunne:2005rt,Lee:2008hz}. Subsequent studies were built upon these results, with notable examples of \cite{Hawking:1981fz,Hawking:1998bn,Hackworth:2004xb}, all of which found new classes of solutions, with the latter two only appearing in the presence of gravity. More recently, Dong and Harlow \cite{Dong:2011gx} presented exact solutions for the CDL bounce in higher dimensions, though such solutions require potentials that do not have simple closed-form expressions. In four dimensions, it is possible to obtain exact bounce solutions if the potential contains exponential terms \cite{Kanno:2011vm}. The particular case where the scalar potential is symmetrical (i.e., with degenerate vacua) was considered in \cite{Lee:2011ms} and provides a possible braneworld interpretation. In \cite{Gregory:2013hja}, it was found that the presence of inhomogeneities enhances the tunnelling rate.

A class of solutions corresponding to multiple bounces was considered by Hackworth and Weinberg \cite{Hackworth:2004xb}, where the field crosses the potential barrier multiple times before settling at either the false or true vacuum as its final state. A study of fluctuations around these higher-bounce solutions shows that they contain more than one negative mode \cite{Dunne:2006bt} and thus thermal transitions dominate over those attributed to quantum tunnelling. Therefore, these higher-bounce solutions are more akin to the Hawking-Moss solutions \cite{Hawking:1981fz}. The nature of the bounce solutions also depends on the size of the bubbles. It turns out that small bubbles behave similarly to its flat spacetime counterparts, and that large bubbles have no flat space analogues \cite{Koehn:2015hga}.

Using a semi-classical approach with numerical integration, we verified claims by \cite{Hackworth:2004xb} and \cite{Dunne:2006bt} that oscillating bounce solutions with more oscillations appear as the barrier curvature increases. We also verified claims by \cite{Dunne:2006bt} that these new oscillating bounce solutions have more than one negative mode, which corresponds to thermalisation. On the other hand, Coleman-De Luccia (CDL) bounce solutions, also referred to as 1-bounce solutions, have only one negative mode and thus corresponds to tunnelling\cite{Dunne:2006bt}.

While doing so, we obtained a class of tunnelling bounce solutions with a non-vanishing scale factor, suggesting that the boundary conditions imposed by \cite{Coleman:1980aw} and subsequent work may be relaxed. Unlike the CDL solutions or the more general oscillating bounce solutions, these cannot be considered a subset of Hawking-Turok solutions due to the non-vanishing scale factor. This suggests that bounce solutions in general are not a subset of Hawking-Turok solutions. In addition, like the CDL bounce solutions, these solutions have only one negative mode, which suggest that it is possible to have quantum tunnelling effects without the scale factor vanishing for finite Euclidean time. This solution was only found for intial conditions beginning near the true vacuum, and not for initial conditions beginning near the false vacuum.

In addition, our results suggest that the bounce solutions found by  \cite{Hackworth:2004xb} and \cite{Dunne:2006bt} may vanish as barrier curvature increases, starting from solutions with the smallest number of oscillations. However there is one particular exception: For solutions corresponding to transitions from true to false vacua, the 1-bounce and 2-bounce solution vanish at the same value of barrier curvature. The 1-bounce solution has a further property where as the barrier curvature increases, it turns into a new class of bounce solution (which we will describe in Sec.~\ref{new}) before vanishing. Instead, it appears to vanish at the value of barrier curvature for which the new class of bounce solution begins, suggesting that the mechanism responsible for the vanishing solutions may be related to the new class of bounce solution.

This paper is organised as follows: In Sec.~\ref{classical} we review the Euclidean action and derive the classical equations of motion for the theory. In the same section we will also provide a description of the potential function used in this paper. Subsequently, we give a brief review of the known solutions in Sec.~\ref{known}. In Sec.~\ref{new} we present a bounce solution that satisfies a different set of boundary conditions from the earlier known solutions. In Sec.~\ref{vanishing}, we provide numerical evidence that the existence of oscillating bounce solutions require that the barrier curvature lie within a certain range of values. This paper concludes with some closing remarks in Sec.~\ref{conclusion}.

\section{Action and equations of motion} \label{classical}

In the semi-classical limit, the decay rate per volume of the meta-stable state is given by \cite{Coleman:1977py,Coleman:1980aw}
\begin{align}
 \Gamma\propto \expo{-S_{\mathrm{E}}},
\end{align}
where $S_{\mathrm{E}}$ is the Euclidean action of a four-dimensional self-interacting scalar field minimally coupled to gravity,
\begin{align}
 S_{\mathrm{E}}&=\int\dif^4x\sqrt{g}\brac{\half\brac{\nabla\phi}^2+V(\phi)-\frac{1}{2\kappa}R},
\end{align}
where $\kappa=8\pi/M_{\mathrm{Pl}}^2$ gives the coupling to gravity via the Ricci scalar $R$. As mentioned in Sec.~\ref{intro}, it is of interest to study the classical solutions to this action, which correspond to $\delta S_{\mathrm{E}}=0$. Varying the action with respect to the scalar field and metric leads to the Einstein-scalar equations
\begin{subequations}\label{EOM}
\begin{align}
 R_{\mu\nu}&=\kappa\brac{\nabla_\mu\phi\nabla_\nu\phi-V g_{\mu\nu}},\label{EE}\\
 \nabla^2\phi&=\frac{\dif V}{\dif\phi}.\label{KGE}
\end{align}
\end{subequations}

Coleman et al.~has shown that bounce solutions are always $O(4)$-symmetric \cite{Coleman:1977th}. Therefore we take our geometry to be described by the metric
\begin{align}
 \dif s^2&=\dif\sigma^2+a(\sigma)^2\dif\Omega_{(3)}^2,\label{metric}
\end{align}
where $\dif\Omega^2_{(3)}$ is the metric of a round $S^3$. With this $O(4)$ symmetry reflected in the geometry, $\phi$ can be taken to depend only on `Euclidean time' $\sigma$, then the equations of motion \Eqref{EOM} reduce to
\begin{subequations} \label{EOM2}
\begin{align}
 \ddot{\phi}&=-3\frac{\dot{a}}{a}\dot{\phi}+\frac{\dif V}{\dif\phi},\label{phiddot}\\
 \ddot{a}&=-\frac{1}{3}\kappa a\brac{\dot{\phi}^2+V},\label{addot}
\end{align}
\end{subequations}
along with a constraint
\begin{align}
 \dot{a}^2+\frac{1}{3}\kappa a^2\brac{V-\half\dot{\phi}^2}=1.\label{constraint}
\end{align}
Here, over-dots denote derivatives with respect to $\sigma$.

\begin{figure}
 \begin{center}
  \includegraphics{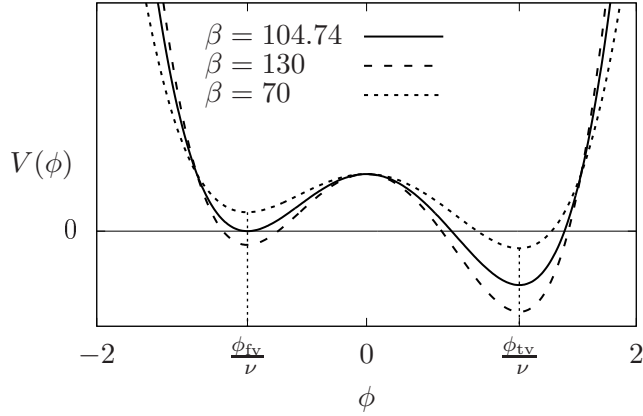}
  \caption{Plots of $V(\phi)$ for several $\beta$ with $\epsilon=0.23$ and $b=0.25$. With these parameters, $\beta=104.74$ corresponds to $V(\phi_{\mathrm{fv}})\simeq0$.}
  \label{fig_Vplot}
 \end{center}
\end{figure}

In the following, we consider a scalar potential of the form
\begin{align}
 V(\phi)=V_{\mathrm{top}}+\beta H^2_{\mathrm{top}}\nu^2\brac{-\half\brac{\frac{\phi}{\nu}}^2-\frac{b}{3}\brac{\frac{\phi}{\nu}}^3+\frac{1}{4}\brac{\frac{\phi}{\nu}}^4}. \label{Potential}
\end{align}
This potential has two local minima, namely the false vacuum $\phi=\phi_{\mathrm{fv}}$ and the true vacuum $\phi=\phi_{\mathrm{tv}}$, where $V(\phi_{\mathrm{tv}})<V(\phi_{\mathrm{fv}})$. The two minima are separated by a barrier at $\phi=0$, as shown in Fig.~\ref{fig_Vplot}. The parameter $b$ sets the spacing between $\phi_{\mathrm{fv}}$ and $\phi_{\mathrm{tv}}$, with $\phi_{\mathrm{fv}}<0<\phi_{\mathrm{tv}}$, while $\nu$ sets the scale for the scalar field, which we may parametrise more conveniently as $\epsilon=\sqrt{\kappa\nu^2/3}$. The parameter $\beta$ is the `barrier curvature', as it is related to the second derivative of the potential at $\phi=0$ by
\begin{align}
 \frac{\dif^2 V}{\dif\phi^2}\bigg|_{\phi=0}=-H_{\mathrm{top}}^2\beta.
\end{align}
For later convenience, we further define
\begin{align}
 H(\phi)=\sqrt{\frac{\kappa}{3}\left|V(\phi)\right|}.
\end{align}
We will also find it convenient to introduce shorthands for the following quantities:
\begin{align}
 H_{\mathrm{fv}}=\sqrt{\frac{\kappa}{3}\left|V(\phi_{\mathrm{fv}})\right|},\quad H_{\mathrm{tv}}=\sqrt{\frac{\kappa}{3}\left|V(\phi_{\mathrm{tv}})\right|},\quad 
 H_{\mathrm{top}}=\sqrt{\frac{\kappa}{3}\left|V_{\mathrm{top}}\right|}. \label{special_H}
\end{align}

Throughout most of this paper, we will be concerned with finding numerical solutions to Eq.~\Eqref{EOM2} under the potential \Eqref{Potential}. To do this we implement the standard shooting method where we begin the integration at $\sigma=0$ with the relevant initial conditions which we will elaborate below. Our shooting parameter will be $\phi(0)$, which we adjust until we find a solution which satisfies the appropriate end-point boundary conditions. The constraint \Eqref{constraint} is used to set the scale of $a(\sigma)$.

\section{Review of known solutions} \label{known}

In this section, we review the various solutions found in earlier literature. Aside from the trivial solutions, these can be categorised by the boundary conditions, namely the behaviour of $\phi$ and $a$ at the boundaries of the domain $0\leq\sigma\leq\sigma_{\mathrm{max}}$ for some positive $\sigma_{\mathrm{max}}$.

\textit{Trivial solutions}: The simplest solutions to Eqs.~\Eqref{phiddot} and \Eqref{addot} are the trivial solutions with $\phi$ being constant at $0$, $\phi_{\mathrm{fv}}$, or $\phi_{\mathrm{tv}}$. The particular case $\phi=0$ is known in the literature as the Hawking-Moss solution \cite{Hawking:1981fz}. With a constant $\phi$, Eq.~\Eqref{phiddot} is trivially satisfied and Eq.~\Eqref{addot} is solved by
\begin{align}
 a(\sigma)=\left\{ \begin{array}{cc}
                          \frac{1}{H(\phi)}\sin\sbrac{H(\phi)\sigma}, & V(\phi)>0, \\
                          \frac{1}{H(\phi)}\sinh\sbrac{H(\phi)\sigma}, & V(\phi)<0,\\
                          \sigma, & V(\phi)=0.
                         \end{array}\right. \label{TrivialSolutions}
\end{align}

\begin{figure}
 \begin{center}
  \includegraphics[scale=0.85]{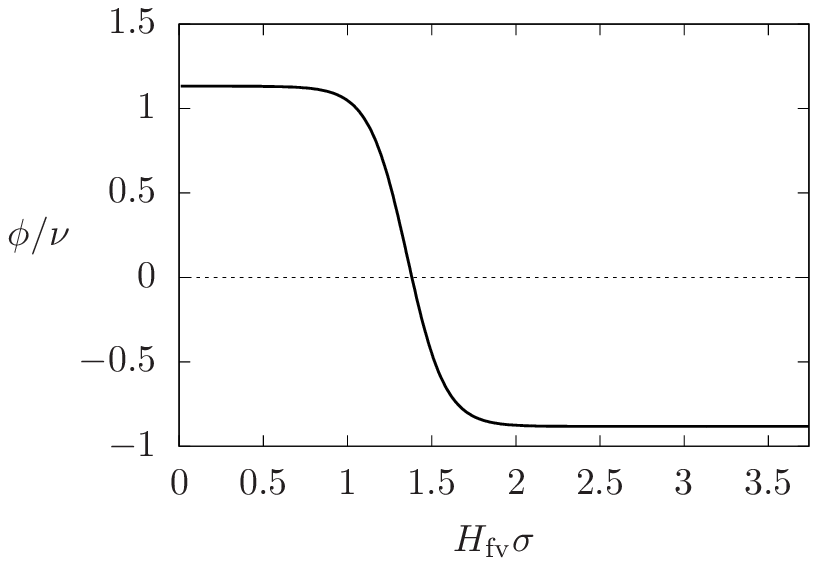}
  \includegraphics[scale=0.85]{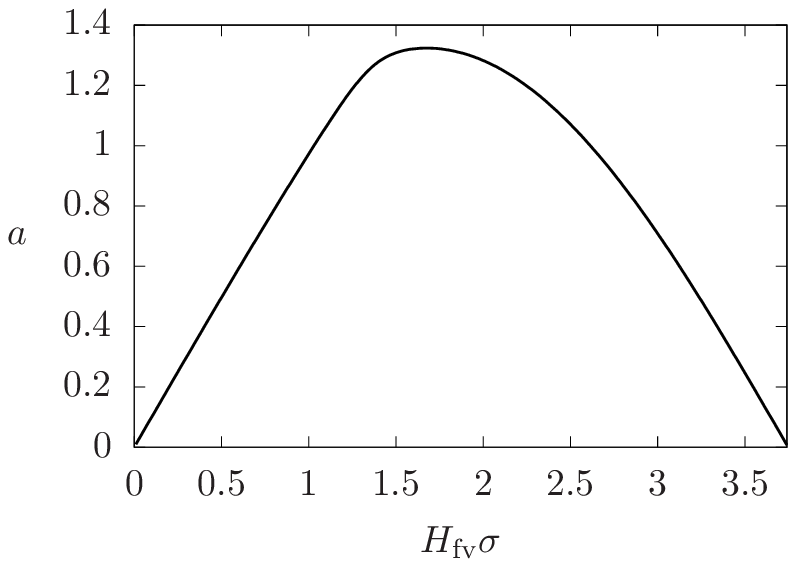}
  \caption{The CDL, or 1-bounce solution with $\beta=45$, $b=0.25$, and $\epsilon=0.23$. The horizontal axis is plotted in units of $H_{\mathrm{fv}}\sigma$, where $H_{\mathrm{fv}}$ is defined in Eq.~\Eqref{special_H}.}
  \label{fig_beta_45_1_bounce}
 \end{center}
\end{figure}

\begin{figure}
 \begin{center}
  \includegraphics[scale=0.85]{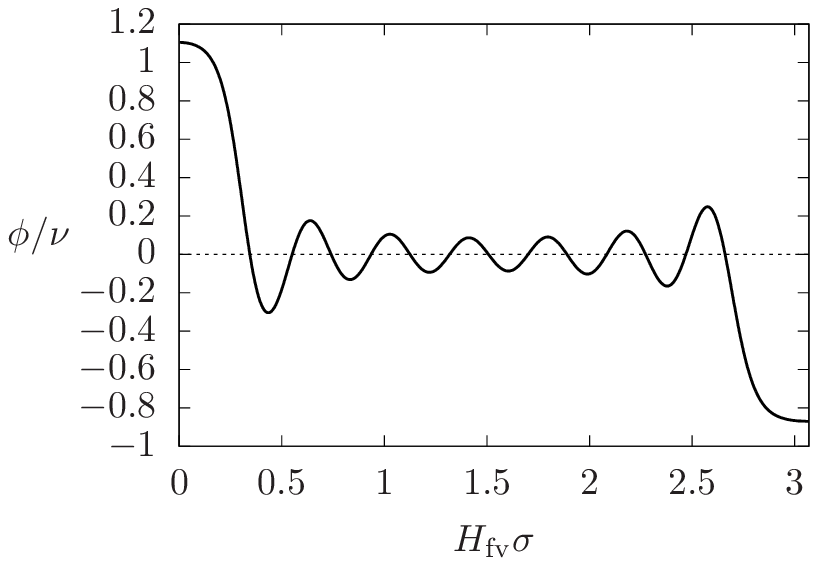}
  \includegraphics[scale=0.85]{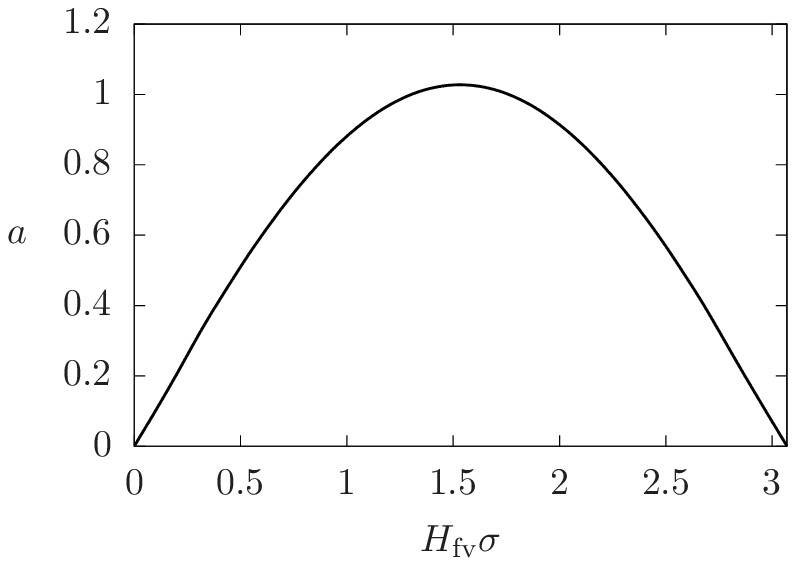}
 \end{center}
 \caption{Example of an oscillating bounce solution with $n=13$ with $\beta=45$, $b=0.25$, and $\epsilon=0.23$. The horizontal axis is plotted in units of $H_{\mathrm{fv}}\sigma$, where $H_{\mathrm{fv}}$ is defined in Eq.~\Eqref{special_H}.}
 \label{fig_MoreBounces}
\end{figure}

\textit{Bounce solutions}: These are solutions defined by the boundary conditions
\begin{align}
 \dot{\phi}(0)&=0,\quad\dot{\phi}(\sigma_{\mathrm{max}})=0,\nonumber\\
 a(0)&=0,\quad a(\sigma_{\mathrm{max}})=0, \label{BC_Bounce}
\end{align}
where the last condition is taken as the defining equation for $\sigma_{\mathrm{max}}$, particularly the point where $a(\sigma)$ returns to zero at some finite $\sigma$. The typical scenario of interest is where $\phi$ starts somewhere close to $\phi_{\mathrm{tv}}$ at $\sigma=0$, and ends near $\phi_{\mathrm{fv}}$ at $\sigma=\sigma_{\mathrm{max}}$.  
Over this interval of $\sigma$, $\phi$ may oscillate, or bounce, multiple times across the barrier at $\phi=0$. A bounce solution can then be characterised by its number of bounces, $n$, where $n$ as the number of times the field crosses the $\phi=0$ barrier.\footnote{Or equivalently, $n$ is the number of roots of the equation $\phi(\sigma)=0$.} In particular, the single-bounce solution depicted in Fig.~\ref{fig_beta_45_1_bounce} was found by Coleman and De Luccia \cite{Coleman:1980aw} (CDL) and we shall refer to it as the `CDL bounce solution', as we have alluded to in Sec.~\ref{intro}. Solutions with more bounces are those considered in \cite{Hackworth:2004xb,Dunne:2006bt}. An example of $n=13$ is shown in Fig.~\ref{fig_MoreBounces}. We shall denote these $n\geq 2$ solutions as `oscillating bounce solutions'.

\textit{Hawking-Turok/ $\phi$-diverging solutions}: The solutions considered by Hawking and Turok \cite{Hawking:1981fz} are those with the boundary conditions
\begin{align}
 \dot{\phi}(0)&=0,\nonumber\\
 a(0)&=0,\quad a(\sigma_{\mathrm{max}})=0. \label{BC_HT}
\end{align}
Between the discrete bounce solutions, there exist a continuous range of solutions where $\phi(\sigma_{\mathrm{max}})\rightarrow\pm\infty$ instead of ending at some finite $\phi(\sigma_{\mathrm{max}})$ with $\dot{\phi}(\sigma_{\mathrm{max}})=0$. We shall also call these $\phi$-diverging solutions, in part to distinguish this class from a separate one we consider in the next Section where $a$ diverges instead of $\phi$. It was suggested by Ref.~\cite{Hawking:1998bn} that these are valid classical solutions, but are ignored because they diverge and are non-discrete. We also observe that Eq.~\Eqref{BC_HT} is simply Eq.~\Eqref{BC_Bounce} with the condition $\dot{\phi}(\sigma_{\mathrm{max}})=0$ relaxed. Hence we may regard the bounce solution as a subset of the Hawking-Turok solutions.

\section{Bounce solutions with diverging scale factor}\label{new}

In addition to the three classes of solutions outlined in the previous section, we report in this paper another class of solutions that, to our knowledge, has yet to be studied in detail. The boundary conditions are
\begin{align}
 \dot{\phi}(0)&=0,\quad\dot{\phi}(\sigma\rightarrow\infty)=0,\nonumber\\
 a(0)&=0. \label{BC_New}
\end{align}
Similar to the Hawking-Turok solutions, these conditions are simply Eq.~\Eqref{BC_Bounce} with one condition relaxed. In the present case, the condition we relax is the requirement that $a(\sigma)$ approaches zero within finite $\sigma$. Nevertheless, our solution here is distinct from the Hawking-Turok in the sense that $\phi$ is always finite, whereas $\phi$ diverges in the Hawking-Turok case. Unlike all the known solutions outlined in the previous sections, the range of $\sigma$ for the present class is semi-infinite, $\sigma\in[0,\infty)$.\footnote{As Eq.~\Eqref{EOM2} are non-linear equations, the number of boundary conditions is not always the same for special solutions. For our present pruposes, it is only crucial that the initial conditions are fixed, and the set of solutions should be discretised. A continuous family of solutions are not helpful in calculating path integrals.} Due to the boundary conditions on $\phi$, the solutions of $\phi$ do not diverge. Therefore it is similar to the bounce solutions outlined in Sec.~\ref{known}, except that $a$ diverges as $\sigma\rightarrow\infty$. Therefore, we shall henceforth refer to this class as the $a$\emph{-diverging solutions}. For these solutions to occur, the potential is required to satisfy
\begin{align}
 V(\phi_{\mathrm{fv}})<0\quad\mbox{ and }\quad V(\phi_{\mathrm{tv}})<0. \label{AdS_to_AdS}
\end{align}

The boundary conditions \Eqref{BC_New} are distinct from those reviewed in Sec.~\ref{known}, thus one might regard this as a new class of solutions. However if viewed in terms of transitions between two AdS vacua (as is clear from Eq.~\Eqref{AdS_to_AdS}), these solutions are implicit in the constructions considered in, among others, Refs.~\cite{Coleman:1980aw,Parke:1982pm,Brown:1988kg}, most of which were calculated under thin-wall approximations. More recently, numerical solutions in the case of Minkowski to AdS vacuum has been studied by Masoumi et al. in \cite{Masoumi:2016pqb}.\footnote{We thank our anonymous referee for pointing this out.} 

\begin{figure}
 \begin{center}
  \includegraphics[scale=0.85]{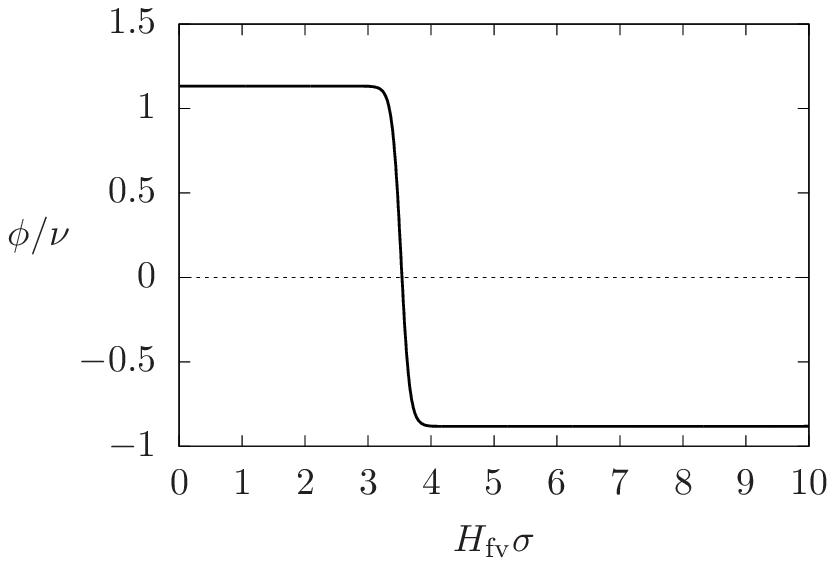}
  \includegraphics[scale=0.85]{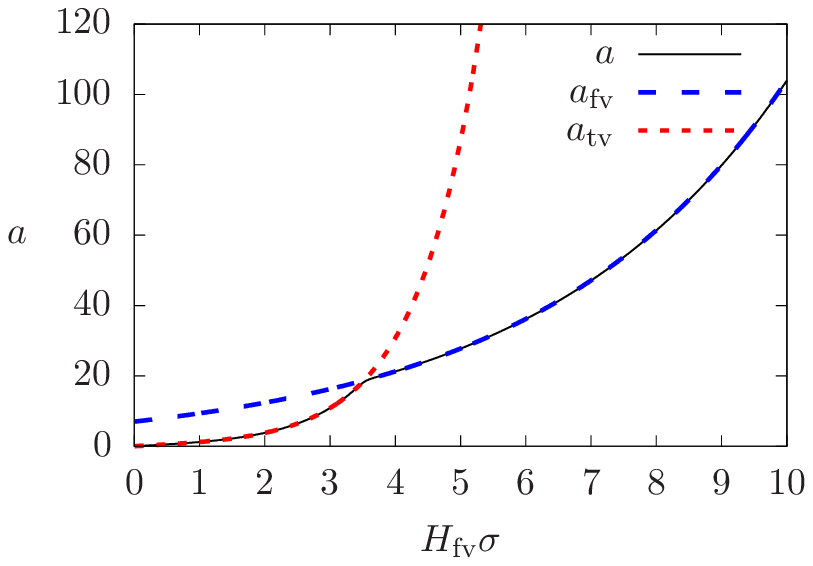}
  \caption{(Colour online) New class of bounce solution, shown here for the case $\beta=112$, $b=0.25$, $\epsilon=0.23$, and $H_{\mathrm{fv}}$ as defined in Eq.~\Eqref{special_H}. For these parameters, the initial value of $\phi$ is $\phi=\phi_{\mathrm{tv}}-3.6855\ldots\times 10^{-21}$. In the plot for $a$ vs $H_{\mathrm{fv}}\sigma$, the dotted and dashed lines are $a_{\mathrm{fv}}$ and $a_{\mathrm{tv}}$ given in Eq.~\Eqref{CompareTrivial}, the corresponding trivial solutions which can be seen to coincide closely with the new solution in their corresponding segments.}
  \label{fig_NewBounce}
 \end{center}
\end{figure}

An example of this $a$-diverging class is shown in Fig.~\ref{fig_NewBounce}, where we can see that it consists of two distinct regions of $\dot{\phi}\simeq0$ and negative $V(\phi)$ connected by a very brief transition. In fact, it spends most of its time near the vacua such that $\phi$ and $a$ behave like their respective trivial solutions. As a result, the scale factor $a$ accordingly has two sinh-like regions, each of which well-approximated by the negative $V(\phi)$ case in \Eqref{TrivialSolutions}. This is shown explicitly in Fig.~\ref{fig_NewBounce} where we have also plotted $a_{\mathrm{fv}}$ and $a_{\mathrm{tv}}$, the trivial solutions given by 
\begin{align}
 a_{\mathrm{fv}}&=\frac{1}{H_{\mathrm{fv}}}\sinh\brac{H_{\mathrm{fv}}\sigma+1.37},\quad
 a_{\mathrm{tv}}=\frac{1}{H_{\mathrm{tv}}}\sinh\brac{H_{\mathrm{tv}}\sigma}, \label{CompareTrivial}
\end{align}
and we have exploited the invariance of the equations of motion under $\sigma\rightarrow\sigma+\mathrm{constant}$ to see that the growth rate coincides with the new solution after the transition.

When $\phi$ is initially near the true vacuum, the scale factor increases exponentially at a rate which is approximately $H\simeq H_{\mathrm{tv}}$. When $\phi$ bounces over to near the false vacuum, $a$ is still exponentially increasing, but at a different rate $H\simeq H_{\mathrm{fv}}$. At present, the $a$-diverging class of solution with the above-mentioned behaviour of $a$ has only been found for single-bounces of $\phi$. We believe that sinh-like scale factors for oscillating bounce solutions are not found because the transition region is too large.

We regard this class also as a bounce solution since the scalar field continues to behave in a similar manner as the CDL bounce solutions. However the domain of this solution is $\sigma\in[0,\infty)$, instead of ending at a finite $\sigma$ as in the CDL case. Clearly, this cannot be considered a subset of the Hawking-Turok solutions, since, as mentioned above, $a$ diverges and thus do not obey Hawking-Turok boundary conditions in Eq.~\Eqref{BC_HT}.

Additionally, we have studied the quantum fluctuation about these classical solutions using a method similar to \cite{Dunne:2006bt}. The $a$-diverging solutions only have one negative mode, which is identical to the results obtained for the CDL solutions, suggesting that this class of solutions also correspond to quantum tunnelling effects \cite{Dunne:2006bt}. This also suggests that the solutions are of a similar vein of bounce solutions with the currently known CDL and oscillating bounce solutions. It further reinforces the fact that bounce solutions, in general, are not subsets of a general Hawking-Turok solution, but rather the choice of which three boundary condition should be used is not restricted as long as we have $\dot{\phi}=0$ at the two end-points.

\section{Criteria of existence for bounce solutions} \label{vanishing}

For the various classes of bounce solutions we described above, we will demonstrate that there is an upper bound of $\beta$ for which bounce solutions can exist. When the value $\beta$ goes beyond this bound, the bounce solutions no longer exist. This results supplements earlier arguments that $\beta$ has a \emph{lower} bound. In particular, Ref.~\cite{Hackworth:2004xb} found that for the $n$-oscillating bounce solution to exist, the value of $\beta$ must have a sufficiently large value, namely
\begin{align}
 \beta>n(n+3). \label{beta_bound}
\end{align}
This result was obtained by considering small perturbations about $\phi_{\mathrm{top}}$, and was verified numerically in the non-perturbative case. This continues to hold true for the non-perturbative oscillating bounce solutions in the large-$n$ limit \cite{Dunne:2006bt}. 

Here, we demonstrate that there also exist an \emph{upper} bound to $\beta$, which we will denote as $\beta_*$, in addition to demonstrating that the non-perturbative, oscillating bounce solutions are consistent with \Eqref{beta_bound}. We show in this section that the oscillating bounce solutions require 
\begin{align}
 n(n+3)<\beta<\beta_*,
\end{align}
for some value $\beta_*$ which we will explore in this section.

In general, we observe that as we increase $\beta$ further away from $n(n+3)$, the initial value $\phi(0)$ required to obtain various bounce solutions approaches the minima $\phi_{\mathrm{tv}}$. In other words, as $\beta$ increases, the difference $\phi_{\mathrm{tv}}-\phi(0)$ for the oscillating bounce solution and the $a$-diverging solution becomes smaller. This difference continues to diminish until a value of $\beta=\beta_*$ where the difference effectively vanishes and only the trivial solution remains. We shall call this upper bound $\beta_*$ the `\textit{vanishing point}'.

\begin{figure}[H]
 \begin{center}
  \includegraphics[scale=0.85]{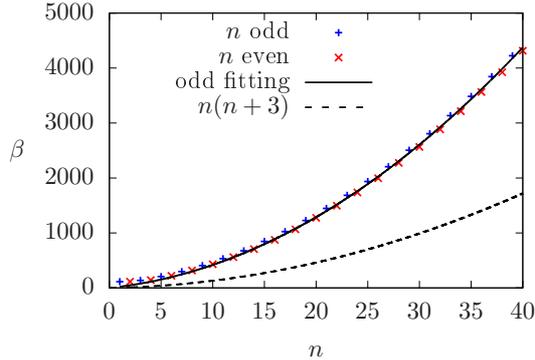}
  \caption{(Colour online) Values of $\beta_*$ for which the bounce solutions vanish, with $b=0.25$, and $\epsilon=0.23$. The solid line is the plot of Eq.~\Eqref{FitsTVFV} for the odd-bounce solutions. The dashed line is the curve $\beta=n(n+3)$, the lower bound estimated by \cite{Hackworth:2004xb}.}
  \label{fig_Vanishing}
 \end{center}
\end{figure}

We first consider solutions corresponding to the true vacuum to false vacuum transitions. We see that the $n=1$ CDL solution vanishes first with the smallest $\beta_*$ compared to the $n=3$ and $n=5$ solutions. Table \ref{tab_VanishingTVFV} shows the vanishing points for bounces up to $n=10$. We note that the $n=1$ (the CDL bounce) and $n=2$ bounce vanishes first, at $\beta_*=112.4$, followed by the vanishing of the higher bounce solutions at larger $\beta_*$. Therefore, beyond a certain $\beta$, there are no longer any quantum transitions due to the CDL bounce, but rather thermal ones with $n>1$. For these values of $\beta_*$, the potentials are negative for both the false and true vacua. The vanishing points for bounces up to $n=40$ are plotted in Fig.~\ref{fig_Vanishing}. 

\begin{table}[H]
 \begin{center}
 \begin{tabular}{rcccccccccc} 
  \hline
   $n$:        & 1 & 2 & 3 & 4 & 5 & 6 & 7 & 8& 9 & 10\\
   $\beta_*$:  & 112.4 & 112.4 & 136.5 & 142.7 & 206.2 & 220.0 & 297.0 & 316.4 & 406.7 & 431.1\\
  \hline
 \end{tabular}
 \caption{The vanishing points of $n$-bounce solutions up to $n=10$ for the true vacuum to false vacuum transitions, with $b=0.25$ and $\epsilon=0.23$.}
 \label{tab_VanishingTVFV}
 \end{center}
\end{table}

It is worth observing in Fig.~\ref{fig_Vanishing} that the odd and even bounces can be considered to lie on separate, slightly different trend-lines, with $\beta_*$ for the odd bounces are always slightly higher than those of the even bounces. We may also estimate a functional relation for each trendline by performing a Richardson extrapolation on the data points. We find that $\beta_*$ has the form
\begin{align}
 \beta_*\sim \frac{9}{4}n(n+\zeta),\quad n\rightarrow\infty, \label{FitsTVFV}
\end{align}
where the odd and even solutions have the parameter $\zeta$ given by
\begin{align}
 \zeta\simeq\left\{\begin{array}{cc}
                8.61, & \mbox{odd }n,\\
                7.62, & \mbox{even }n.
               \end{array}\right.               
\end{align}
This is one of the major results of this paper. The fitting estimated in \Eqref{FitsTVFV} for $\zeta=8.61$ (odd $n$) is plotted as the solid line in Fig.~\ref{fig_Vanishing}. The even-$n$ case is not shown as it is nearly indistinguishible from the odd-$n$ plot on a figure of that scale. Furthermore, we have verified that all the bounce solutions satisfies the condition \Eqref{beta_bound} given by \cite{Hackworth:2004xb} under the thin-wall approximation.

We can also perform a similar analysis for the false to true vacuum transitions. The vanishing points of the bounce solutions up to $n=10$ are shown in Table \ref{tab_VanishingFVTV}. It is interesting to note that in this case, the vanishing point for the CDL $n=1$ bounce is $\beta_*=100.63<104.74$. For this value of $\beta$, the potential $V$ is positive at $\phi=\phi_{\mathrm{fv}}$. The values of $\beta_*$ for the higher ($n>1$) bounces all correspond to negative $V$ at both vacua. Performing the Richardson extrapolation for the false-to-true transitions, we find
\begin{align}
 \beta_*\sim \frac{9}{4}n(n+\lambda),\quad n\rightarrow\infty, \label{FitsFVTV}
\end{align}
where 
\begin{align}
  \lambda=\left\{\begin{array}{cc}
                8.667, & \mbox{odd }n,\\
                9.778, & \mbox{even }n.
               \end{array}\right.               
\end{align}

\begin{table}[H]
 \begin{center}
 \begin{tabular}{rcccccccccc} 
  \hline
   $n$:      & 1 & 2 & 3 & 4 & 5 & 6 & 7 & 8 & 9 & 10\\
   $\beta_*$:  & 100.63 & 131.71 & 136.5 & 193.9 & 206.2 & 280.1 & 297.0 & 383.6 & 406.7 & 507.2\\
  \hline
 \end{tabular}
 \caption{The vanishing points of $n$-bounce solutions up to $n=10$ for the false vacuum to true vacuum transitions, with $b=0.25$ and $\epsilon=0.23$ \cite{Dunne:2006bt}.}
 \label{tab_VanishingFVTV}
 \end{center}
\end{table}

As suggested by \cite{Hackworth:2004xb}, the only bounce solution corresponding to quantum tunnelling is the $n=1$ CDL bounce, as the higher-bounce solutions have much larger actions, thus does not contribute greatly to the tunnelling effect. Furthermore, the higher-bounce solutions are those that approach the Hawking-Moss solution, which is a transition that is thermal in nature.

\section{Conclusion} \label{conclusion}

By exploring the numerical solutions to the classical Euclidean equations of motion, we have explored a class of bounce solutions with a diverging scale factor. The behaviour of the scalar field in this $a$-diverging class is similar to that of the CDL single-bounce solution, though the scale factor $a$ grows monotonically with Euclidean time.

The reason for a diverging $a$ is mainly due to this class of solution having vacua corresponding to $V<0$, or a negative effective cosmological constant. Thus the model actually describes false vacuum decay from one anti-de Sitter space into another. While this may not be particularly relevant to inflation or cosmology, it might have potentially interesting holographic descriptions in the context of the AdS/CFT correspondence \cite{Maldacena:2010un,Garriga:2010fu,Kanno:2011hs}. 

We have also numerically verified an argument made in Hackworth and Weinberg \cite{Hackworth:2004xb} that all bounce solutions must satisfy the condition $\beta>n(n+3)$. This argument was originally made based on the thin-wall approximation. Here, our numerical solutions which are found beyond the thin-wall approximation continues to satisfy this bound. In addition, as the barrier curvature $\beta$ increases, the single-bounce CDL solution ceases to exist. Therefore we have provided an additional upper bound and the range of $\beta$ is now bounded to $n(n+3)<\beta<\beta_*$. Since this is the solution with the negative mode that is the main contributor to tunnelling, that means that beyond this value of $\beta$, the remaining modes are those coming from the higher-bounce solutions which behave more similarly to Hawking-Moss thermal transitions.

\bibliographystyle{fvd}

\bibliography{fvd}

\end{document}